\documentclass[rnote]{aa} 
\usepackage{epsfig}
\usepackage{rotating}
\usepackage{natbib}
\usepackage{graphicx}
\newcommand{\camn}{{\em Camera for Multiconjugate Adaptive Optics}}
\newcommand{\cam}{{\em CAMCAO}}
\newcommand{\madn}{{\em Multi-Conjugate Adaptive  Optics Demonstrator}}
\newcommand{\mad}{{\em MAD}}

\newcommand{\vlt}{{\em VLT}}
\newcommand{\vltn}{{\em Very Large Telescope}}
\newcommand{\isaac}{{\em ISAAC}}
\newcommand{\chan}{{\em Chandra}}
\newcommand{\xmm}{{\em XMM-Newton}}

\newcommand{\tmass}{{\em 2MASS}}
\def \oneight{RX J1856.5$-$3754}

\def \zerofour{RX J0420.0$-$5022}

\def \oneseven{1RXS\, J214303.7+065419}

\begin{document}
%

\title{Near infrared \vlt/\mad\ observations of the isolated neutron stars \zerofour\ and \oneight\ \thanks{Based on observations collected at ESO, Paranal, under Program 71.D-0560}}

 \author{R. P. Mignani\inst{1}
 \and
 R. Falomo\inst{2}
\and
A. Moretti\inst{2}
\and
A. Treves\inst{3}
\and 	
 R. Turolla\inst{4,1}
\and
N. Sartore\inst{3}
\and
S.Zane\inst{1}
\and
C. Arcidiacono\inst{2}
\and
M. Lombini\inst{5}
\and
J. Farinato\inst{2}
\and
A. Baruffolo\inst{2}
\and
R. Ragazzoni\inst{2}
\and
E. Marchetti\inst{6}
}

   \institute{Mullard Space Science Laboratory, University College London, Holmbury St. Mary, Dorking, Surrey, RH5 6NT, UK\\
              \email{rm2@mssl.ucl.ac.uk}
\and
INAF, Osservatorio Astronomico di Padova, Vicolo dell'Osservatorio 5, Padua, 35122, Italy 
\and
Department of Physics and Mathematics, University of Insubria, Via Valleggio 11, Como, 22100, Italy
\and
Department of Physics, University of Padua, via Marzolo 8,  Padua, 35131, Italy 
\and
Osservatorio Astronomico di Bologna, INAF, via Ranzani 1, Bologna, 40127, Italy
\and
European Southern Observatory, Karl-Schwarzschild-Strasse 2, Garching, 85748, Germany
}

\titlerunning{Near IR observations of isolated neutron stars}

\authorrunning{Mignani et al.}
\offprints{R. P. Mignani; rm2@mssl.ucl.ac.uk}

\date{Received ...; accepted ...}
\abstract{High-energy observations  have unveiled peculiar  classes of
isolated  neutron stars  which, at  variance with  radio  pulsars, are
mostly radio silent and not powered by the star rotation.  Among these
objects   are   the    magnetars,   hyper-magnetized   neutron   stars
characterized  by transient  X-ray/$\gamma$-ray emission,  and neutron
stars  with  purely  thermal,  and  in most  cases  stationary,  X-ray
emission (a.k.a., X-ray dim  isolated neutron stars or XDINSs).  While
apparently  dissimilar in  their  high-energy behavior  and age,  both
magnetars and XDINSs have  similar periods and unusually high magnetic
fields.  This suggests a  tantalizing scenario where the former evolve
into  the  latter.}{Discovering  so  far  uninvestigated  similarities
between the multi-wavelength properties  of these two classes would be
a further step forward to  establish an evolutionary scenario.  A most
promising channels is the near infrared (NIR) one, where magnetars are
characterized by a distinctive spectral flattening with respect to the
extrapolation of the soft  X-ray spectrum.}{We observed the two XDINSs
\zerofour\ and \oneight\ with the  \madn\ (\mad) at the \vltn\ (\vlt),
as  part of the  instrument guaranteed  time observations  program, to
search for  their NIR  counterparts.}  {Both \oneight\  and \zerofour\
were  not  detected  down  to   K$_s  \sim20$  and  K$_s  \sim  21.5$,
respectively.  }   {In order to constrain the  relation between XDINSs
and  magnetars  it  would  be  of importance  to  perform  deeper  NIR
observations.  A good candidate is  \oneseven\ which is the XDINS with
the highest inferred magnetic field.}


   \maketitle

\section{Introduction}

Over the last two  decades, high-energy observations have unveiled the
existence of  peculiar classes of  isolated neutron stars  (INSs) like
the anomalous X-ray pulsars  and the soft $\gamma$-repeaters (AXPs and
SGRs),  i.e., the magnetar  candidates (e.g.,  \citealt{mererev}), and
the X-ray  Dim INSs (XDINSs; e.g.,  \citealt{frank07}).  Despite their
vastly  different  observational  manifestations, both  magnetars  and
XDINSs  are slow  rotators ($P\sim  1$--10 s),  with  surface magnetic
fields reaching $\approx  10^{14}$--$10^{15}$~G in the magnetars case,
have  persistent  X-ray luminosities  much  larger  than the  inferred
rotational  energy losses $L_{\mathrm  X}\approx 10$--$100\,  \dot E$,
and they are usually radio-silent.

The XDINSs X-ray spectra are  purely thermal and well represented by a
blackbody ($kT\approx 50$--100 eV) whose emission radius is consistent
with a sizable  fraction of the neutron star  surface.  In most cases,
one     or     possibly     more     broad     absorption     features
($E_{\mathrm{line}}\approx  0.2$--0.7  keV)  have been  observed  (see
Haberl 2007 and references therein), which hint toward magnetic fields
of  $\sim 10^{13}-10^{14}$ ~G,  comparable to  those derived  from the
spin  down.  The  absence of  power-law tails  is consistent  with the
faint  rotation--powered  magnetospheric  emission expected  from  the
inferred spin-down luminosity ($\dot E \approx 3-5 \times 10^{30}$ erg
s$^{-1}$).  In  the optical,  only a few  XDINSs have  been identified
(see Kaplan  2008 and references therein) and  a candidate counterpart
was recently proposed  for \oneseven\ (Zane et al.   2008), while that
of  \zerofour\ was  not confirmed  (Mignani et  al.,  in preparation).
Their spectra mostly follow a Rayleigh-Jeans, with fluxes exceeding by
a  factor  $\ge  10$   the  low-energy  extrapolation  of  the  X-rays
best-fitting blackbody \cite[e.g.,][]{kap08}.   Whether this is indeed
due  to  emission  from  regions  of the  star  surface  at  different
temperature \cite[e.g.,][]{pons02} is still under debate.  Like in the
X-rays, no rotation--powered  magnetospheric emission is detectable in
the optical, unless XDINSs are a factor of $\ge 10^{3}$ more efficient
emitters than rotation--powered neutron stars (Zharikov et al.  2006).

Whether there is a  (evolutionary?)  link between magnetars and XDINSs
is still hotly debated (see McLaughlin et al. 2006; Popov et al. 2006,
for  a  discussion). As  proposed  by  Mignani  et al.   (2007a),  the
detection of  the optical/near  infrared (NIR) flattening  observed in
the  magnetar  spectra,  ascribed  either to  magnetospheric  emission
powered by the  magnetic field or to thermal  emission from a fallback
disk, would strengthen  such a link.  For the  XDINSs, such flattening
would be easily recognizable in the NIR, where the contribution of the
optical   Rayleigh-Jeans  continuum  is   negligible.   So   far,  NIR
observations of XDINSs performed with the \vltn\ (\vlt) did not unveil
counterparts down to H= 21.5--22.9 (Lo Curto et al. 2007).

We report on new NIR observations of \oneight\ and \zerofour\ recently
performed in  the K$_s$  band with the  \vlt, the first  presented for
these sources.   Observations are described in Sect.  2, while results
are presented and discussed in Sects. 3 and 4, respectively.

\section{Observations}

\subsection{Observations description}

We performed  K$_s$-band observations  of \oneight\ and  \zerofour\ on
September 28  and 29 2007  (MJD=54371 and 54372),  respectively, using
the      European      Southern      Observatory     (ESO)      \madn\
(\mad)\footnote{http://www.eso.org/projects/aot/mad/},  mounted at the
\vltn\  (\vlt)  Melipal telescope,  as  part  of  the guaranteed  time
program.   The  \mad\ instrument  was  developed  within the  European
Extremely Large Telescope  framework.  The instrument accommodates two
wavefront sensors  (WFS): a  star oriented (SO)  multi Shack--Hartmann
and a layer oriented (LO)  multi pyramid (Regazzoni 2000; Regazzoni et
al. 1996; 2000).   Both sensors can use reference  stars identified in
the  instrument 2\arcmin\  field  of view.   The  \mad\ instrument  is
completed by the \camn\ (\cam), a NIR camera equipped with a $2K\times
2K$ Hawaii  II infrared detector movable over  the 2\arcmin\ corrected
circular field-of-view.  The  pixel-scale is 0\farcs028, corresponding
to  a  57\arcsec$\times$57\arcsec\   field-of-view  on  the  detector.
Thanks to  its flexibility,  the instrument can  be used  in classical
single reference  adaptive optics  (AO), ground layer  adaptive optics
(GLAO),  or multi-conjugate  adaptive optics  (MCAO) according  to the
availability of suitable reference stars close to pointing direction.

During the first night, we used  the LO WFS in classical AO mode, with
a $V\approx 11.3$ reference star. During the second night, we used the
MCAO mode  with 4 fainter stars  as a reference  ($V=14.9, 15.3, 15.4$
and  15.6).   The  seeing  conditions  during  the  observations  were
1\farcs23 for \zerofour\ and  1\farcs37\ for \oneight\ and the airmass
was 1.11 and  1.35, respectively.  For \zerofour, we  obtained a total
of  dithered 65 exposures  along a  pattern that  covered a  region of
$\sim 25\arcsec \times 25\arcsec$.  The exposures were taken in blocks
of  5, i.e.,  along each  node  of the  dithering pattern,  with 30  s
detector  integration time  (DIT) and  NDIT=2.  The  total integration
time  was 3900~s.   WE performed  the  observation of  \oneight\ in  a
similar  way, with  a dithering  pattern  covering a  region of  $\sim
35\arcsec  \times 15\arcsec$. A  total of  25 exposures  were obtained
with DIT=30 s and NDIT=4, for a total integration time of 3000~s.

\subsection{Data reduction and calibration}

For  each set  of science  images we  applied standard  data reduction
procedures for NIR observations, which include trimming of the frames,
bad pixel masking, dark and flat--field correction, and sky-background
subtraction.  Special care was  taken for images alignment, allowing a
correction for rotation to account for a non-optimal efficiency of the
de-rotator.   For the flat--field  correction, we  used the  median of
several images  obtained on  the sky at  the beginning of  each night.
The  bad  pixel  mask  was  created  by analyzing  the  ratio  of  two
flat--field images with  significantly different exposure levels.  For
each science  observation we then  constructed a reference  sky image.
Then we  constructed the  reference sky image  from the median  of all
science images of a given observation. In the case of our targets, the
fields are not  affected by crowding of stellar  sources, which allows
for  the  creation  of  a  good  quality sky  image.   This  was  then
normalized to the median counts  of each science image and subtracted.
Finally, we combined all  sky-subtracted science images after properly
aligning them using as a reference the positions of the brightest (not
saturated) sources detected in all images.

\begin{figure*}
\centering 
 \includegraphics[bb=143 355 473 685,height=8cm,angle=0,clip]{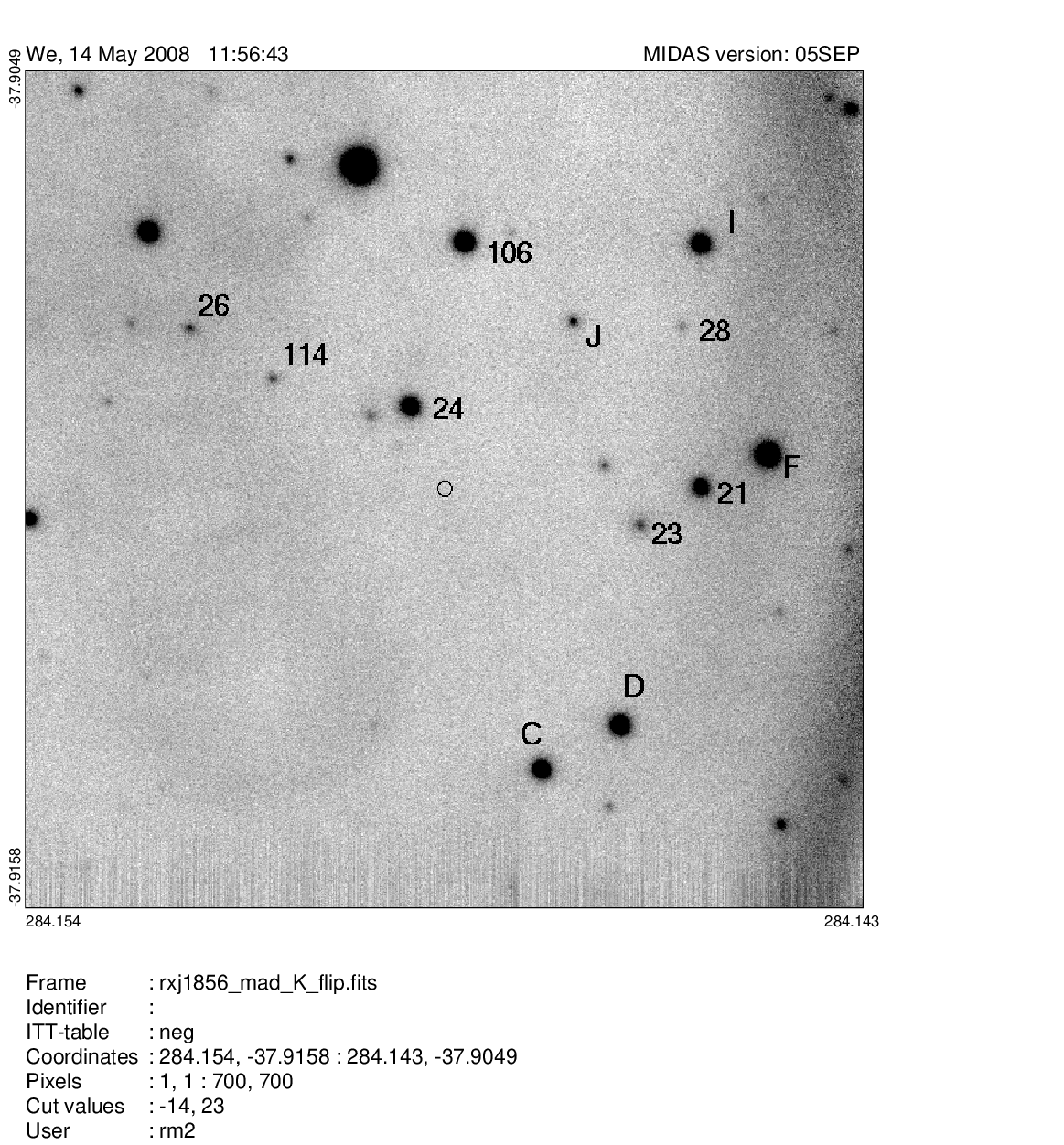}
 \includegraphics[bb=100 300 495 685,height=8cm,angle=0,clip]{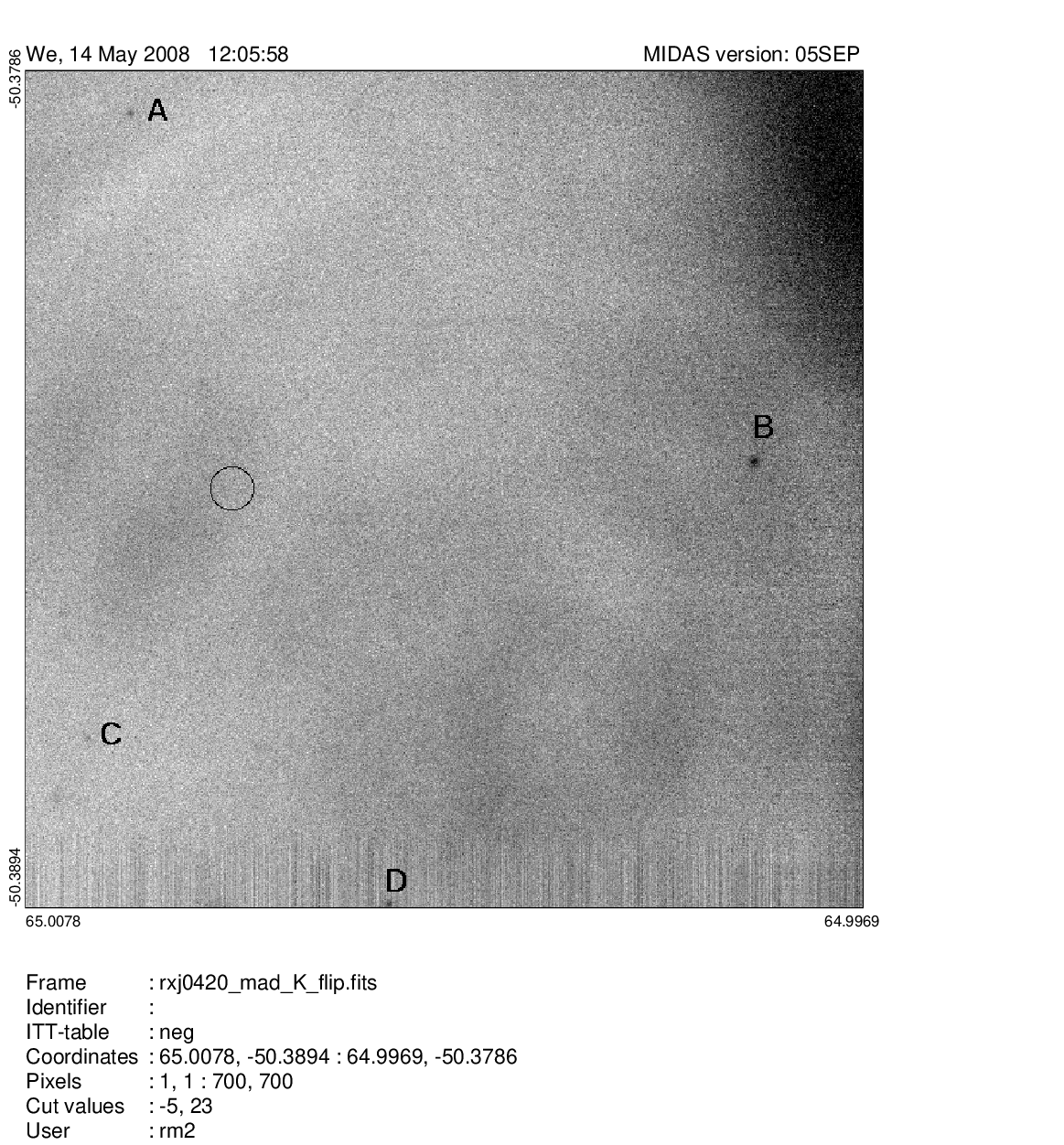}
  \caption{We present $30\arcsec  \times 30 \arcsec$  \mad\ K$_s$-band  images of
 the \oneight\ (left) and \zerofour\ (right) fields. North to the top,
 East  to the  left. The  circles mark the expected  target
 positions, with the  radii  (0\farcs35  and
 1\arcsec,  respectively)  corresponding to their positional uncertainty  (see Sect. 3.1).  The labeling in the \oneight\ field is the
 same used in van Kerkwijk  \& Kulkarni (2001).  The few faint objects
 detected  in the  \zerofour\ field  are also  labeled.  The intensity
 scale in the image has  been adjusted for a better visualization. The
 edge of  the bright  field galaxy  ESO 202-8 is  visible in  the top
 right of the image.  }
\label{fc}       
\end{figure*}

From  the  final co-added  images  we  measured  the associated  image
quality  from the average  full width  at half  maximum (FWHM)  of the
intensity profile of a number of reference stars. Although the nominal
seeing conditions in the night were not optimal (see above), thanks to
the \mad\  AO correction, the FWHM  of stellar images  measured on the
detector actually  turned out  to be 0\farcs28  and 0\farcs40  for the
\zerofour\ and  for the  \oneight\ field, respectively.   We performed
the photometric calibration by observing NIR standard stars during the
same  nights. We  cross-checked our  calibration by  direct comparison
with  objects  of  the  \tmass\  point-source  catalog  (Skrutskie  et
al. 2006).

\section{Data analysis and results}

\subsection{Astrometry}

The astrometry of the \mad\  images of \oneight\ was computed using as
a reference the coordinates and position of 8 \tmass\ stars (Skrutskie
et al. 2006),  all non-saturated and evenly distributed  in the field.
Their  pixel  coordinates were  measured  by  fitting their  intensity
profiles with  a Gaussian function using the  {\em Graphical Astronomy
and           Image          Analysis}           ({\em          GAIA})
interface\footnote{star-www.dur.ac.uk/~pdraper/gaia/gaia.html}.     The
coordinate  transformation  between  the  detector and  the  celestial
reference  frame was then  computed using  the {\em  Starlink} package
{\tt ASTROM}\footnote{http://star-www.rl.ac.uk/Software/software.htm}.
The rms  of the astrometric  solution was $\approx$  0\farcs17.  After
accounting for a 0\farcs2 conservative astrometric accuracy of \tmass,
and the $\sim$ 15 mas accuracy  (Skrutskie et al.  2006) on the tie to
the International Celestial Reference  Frame (ICRF), the overall error
to  be  attached  to  our  astrometry was  finally  0\farcs26.   As  a
reference  for the  \oneight\ coordinates  at the  epoch of  the \mad\
observation  (2007.74)  we  used  those  of  its  optical  counterpart
reported in  Kaplan et  al. (2002), epoch  1999.26, corrected  for the
measured proper motion.  The uncertainty to be attached to the 2007.74
coordinates  includes the  error  on the  proper motion  extrapolation
between  the two  epochs, $\sim  9$ mas  both in  right  ascension and
declination,  and the  error  on the  reference  coordinates which  is
0\farcs2  relative  to  the   ICRF.   We  then  computed  the  overall
uncertainty on  the position determination  of \oneight\ on  the \mad\
images  by   adding  in  quadrature  the  error   on  our  astrometric
calibration  and  the uncertainty  on  its coordinates  extrapolation,
which is $\sim 0\farcs35$.

Since  no  \tmass\  star  is  identified  in  the  much  less  crowded
\zerofour\  field,  we performed  a  two-step astrometric  calibration
using as  a reference a grid  of stars selected from  the \vlt\ H-band
\isaac\ image of  the field (Lo Curto et al.   2007).  By applying the
same procedure  as before, the astrometric calibration  of the \isaac\
image was performed using 8  \tmass\ stars as a reference, yielding an
rms of $\sim 0\farcs22$ on the astrometric solution.  We then computed
the coordinates  of 7  field stars in  common between the  \isaac\ and
\mad\  fields,  which  made   our  secondary  astrometric  grid.   The
astrometric solution yields an rms  of $\sim 0\farcs09$.  By adding in
quadrature all  the errors  of our astrometric  chain, we then  end up
with an accuracy  of 0\farcs32 on our astrometry.   As a reference for
the \zerofour\  coordinates, we used the \chan\  coordinates of Haberl
et al. (2004), which have a nominal error of 0\farcs6 (90\% confidence
level).  We  accounted for the  recently measured proper  motion upper
limit (138 mas yr$^{-1}$; Motch  et al. 2007) when we extrapolated the
\chan\ coordinates of  \zerofour\ (epoch 2002.86) to the  epoch of the
\mad\ observation  (2007.74).  This yields a  position uncertainty due
to the proper motion of  $\sim 0\farcs7$.  By adding in quadrature the
error on  our astrometric  calibration and the  error on  the original
\chan\ coordinates we ended up  with an uncertainty of $\sim 1\arcsec$
on the \zerofour\ position at the epochs of the \mad\ observations.

\begin{figure*}[t]
\centering           
\includegraphics[bb=10 190 440 610,width=8.0cm,angle=0,clip=]{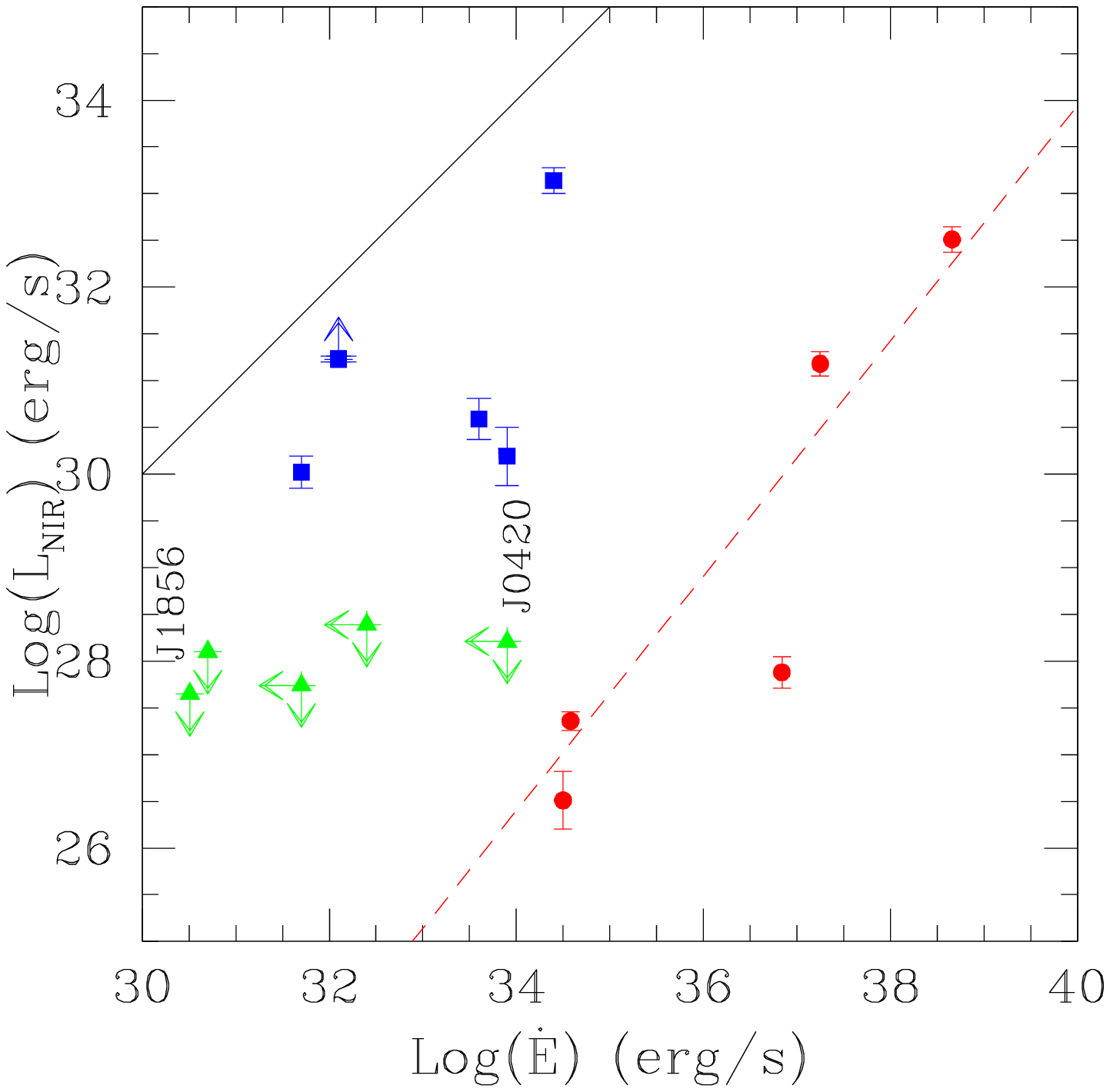} 
\includegraphics[bb=10 190 440 610,width=8.0cm,angle=0,clip=]{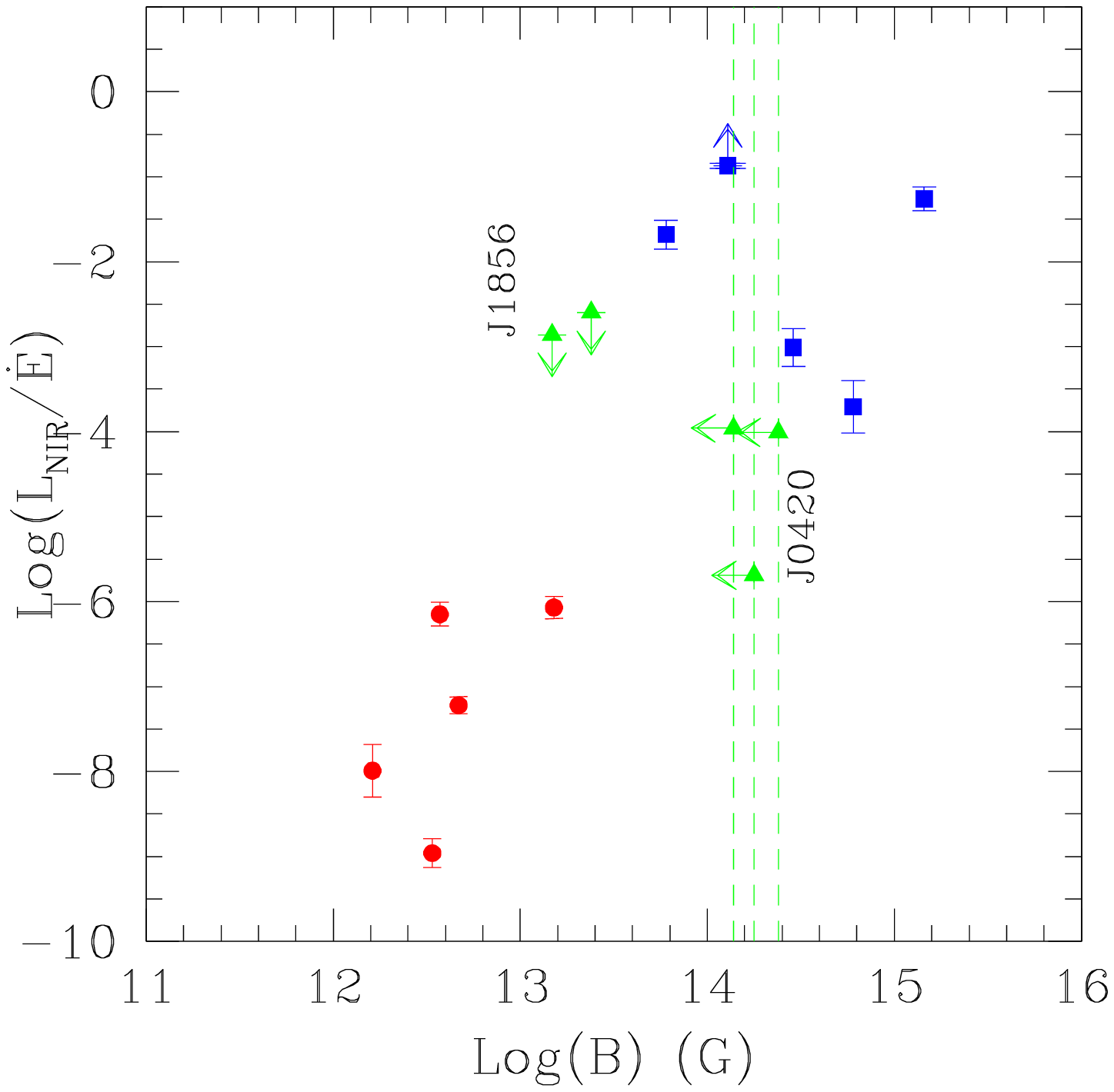}
\caption{(left) INS NIR luminosities  vs. rotational energy loss $\dot
E$.    Filled   circles,   squares   and   triangles   correspond   to
rotation--powered neutron stars,  magnetars, and XDINSs, respectively.
Plots are  updated from Mignani  et al.  (2007b)  and Lo Curto  et al.
(2007), to  which we refer  for the complete  source list and  for the
timing  parameters. Luminosities are  computed in  the K$_s$  band for
magnetars and for rotation--powered neutron stars.  For the XDINSs, we
have  plotted the  luminosity upper  limits computed  from  the H-band
observations of Lo Curto et al.~ (2007).  For \zerofour\ only, we have
plotted  the slightly deeper  luminosity upper  limit obtained from our
K$_s$-band observations.   The red  dashed line is  the linear  fit to
rotation--powered neutron stars, while  the solid line marks the limit
case Log$(L_{NIR})  =$ Log$(\dot  E)$.  (right) INS  NIR efficiencies,
$\eta_{NIR}  \equiv L_{NIR}/\dot  E$, vs.   dipole magnetic  field $B$
inferred from  the spin  down.  Note that  for those XDINSs  for which
only an upper limit on $\dot E$ is available, including \zerofour, the
NIR  emission efficiency is  unconstrained, as  shown by  the vertical
dashed lines.  }
\label{Lir}
\end{figure*}

The two  panels in Fig. 1  show a zoom  of the \mad\ image  around the
expected position of \oneight\  and \zerofour.  Despite of the sharper
PSF of \mad,  our images do not unveil any new  object with respect to
the ones identified  in the long-exposure \isaac\ H-band  images of Lo
Curto  et al.   (2007).  In  both cases,  no candidate  counterpart is
detected at  the expected position,  with the closest  objects located
several arcseconds away.

\subsection{Photometry}

To determine whether our observations  can be used to better constrain
the NIR  emission properties of \oneight\ and  \zerofour, we estimated
their K$_s$-band brightness upper  limits from the limiting magnitudes
of the  \mad\ images.  To  this aim, we  have used as a  reference the
flux of a  number of artificial stars simulated  with the template PSF
measured on the images.  Using  different test values for the PSF flux
normalization,  artificial  stars  were  simulated and  added  to  the
corresponding  images  at  different  coordinates,  randomly  selected
within a few arcsec from the  target position. We then ran in loop the
{\tt Daophot}  detection tool to determine the  PSF flux normalization
corresponding to a $3 \sigma$ detection limit.  In all cases, aperture
correction was applied.  We thus determined that $3 \sigma$ detections
would correspond to magnitudes of  K$_s\sim 21.5$ and K$_s\sim 20$ for
\zerofour\ and  \oneight, respectively.   As an independent  check, we
have  compared  these values  with  the  $3  \sigma$ detection  limits
extrapolated from the  fluxes of the faintest objects  detected in our
images. As  a safe measure,  we used the  \isaac\ H-band images  of Lo
Curto  et  al.   (2007)  to  confirm that  they  were  indeed  genuine
detections and not  background fluctuations.  Following this empirical
approach,  we derived  detection limits  consistent with  the previous
ones.  The  different limiting magnitudes for the  two images, despite
comparable  integration  times,  has  to  be ascribed  mainly  to  the
different image quality between the  two frames, being much better for
\zerofour\ than  for \oneight\  (0\farcs28 with respect  to 0\farcs40,
see  Sect.  2.2).   This  is  indeed a  significant  factor  when  the
signal--to--noise   ratio  is   background  limited.    Moreover,  the
\oneight\  image is  affected by  a higher  background level  which is
$\sim$ 25 \% higher around the target position.

\section{Discussion}

Our measured K$_s$-band spectral flux  upper limits are well above the
extrapolation of  the optical spectrum  of \oneight\ (van  Kerkwijk \&
Kulkarni  2001),  which  would   predict  K$_s  \sim  26.5$,  and  the
Rayleigh-Jeans extrapolation  in the  optical/NIR domain of  the \xmm\
spectrum  of \zerofour\  (Haberl  et  al.  2004).   We  note that,  for
\oneight, the  K$_s$-band spectral flux upper limit  is shallower than
the H-band one  of Lo Curto et al.~(2007), while  for \zerofour\ it is
slightly  deeper. Thus,  we  are not  yet  able to  constrain the  NIR
spectrum of \oneight\ and \zerofour.

A blackbody spectrum produced by the neutron star surface is obviously
a  possibility.  It  is clear  that for  the \oneight\  and \zerofour\
distances of  $\sim 160$  pc (van Kerkwijk  \& Kaplan 2007)  and $\sim
350$ pc (Posselt  et al.  2007), respectively, and  for any reasonable
combination of temperature and emitting area, the predicted flux would
fall well below our upper  limits.  A blackbody spectrum produced by a
fallback disk  (Perna et  al. 2000) could  yield to a  flux compatible
with the  present NIR  flux upper limits  of \oneight\  and \zerofour,
depending  on the actual  disk size  and accretion  rate (see  also Lo
Curto et al.  2007).
A power--law spectrum could  be produced by non--thermal emission from
the neutron  star magnetosphere.   In Fig.\ref{Lir} (left),  we plotted
the NIR  luminosity ($L_{NIR}$) vs  the rotational energy  loss ($\dot
E$) for  different classes  of INSs: rotation--powered  neutron stars,
magnetars, and XDINSs. Our upper limits do not rule out a NIR emission
efficiency, $\eta_{NIR} \equiv L_{NIR}/\dot  E$, comparable to that of
rotation--powered neutron  stars, or higher.  This  would be possible,
for instance,  if the fraction  of energy emitted in  the relativistic
wind is  much less than in  rotation--powered neutron stars  or if the
NIR emission is powered not by the neutron star rotation but, e.g.,  by
its magnetic field.  Although  XDINSs are less luminous than magnetars
in absolute terms (Fig.\ref{Lir},  left), they might have a comparable
NIR emission efficiency.  This is shown in Fig.\ref{Lir} (right) where
the NIR emission efficiency is  plotted vs.  the dipole magnetic field
$B$ inferred  from the  spin down.  
We  warn  here   that  NIR  observations  of  the   radio  pulsar  PSR
J1119$-$6127 (Mignani et al.  2007b) suggest a NIR emission efficiency
comparable to  that of other rotation--powered  neutron stars, despite
of its relatively high magnetic field ($B \sim 4.1 \times 10^{13}$ G).
If the inferred  magnetic field value of PSR  J1119$-$6127 is correct,
this  might  argue  against  a  connection between  the  NIR  emission
efficiency and the magnetic field.

Much deeper  observations are required  to better constrain  the XDINS
optical/NIR  emission  properties and  to  investigate their  possible
connection  with   the  magnetars.   In   this  respect,  particularly
important would be the study of \oneseven, possibly the XDINS with the
highest  magnetic field,  as  hinted  by the  detection  of the  X-ray
absorption feature (Zane et  al.  2005).  Its anomalously high optical
emission is  indeed incompatible with  both thermal emission  from the
neutron star surface and rotation--powered non--thermal emission (Zane
et al. 2008).

\begin{acknowledgements}
RPM acknowledges STFC for  support through a rolling grant
and  the ESO/Chile Visitors  Program for hospitality at the Santiago Office  where most of this work
was finalized.  RPM also  thanks P.  Amico and G.  Lo Curto  (ESO) for
useful  discussions.   SZ acknowledges  STFC  for  support through  an
Advanced Fellowship.

\end{acknowledgements}

\end{document}